\documentclass[acmtog,screen]{acmart}

\usepackage{pgfplots}
\usepackage{pgfplotstable}
\usepgfplotslibrary{groupplots}
\pgfplotsset{
    width=0.3\linewidth,
    compat=1.9,
}

\usepackage{stfloats}

\copyrightyear{2026}
\acmYear{2026}
\setcopyright{cc}
\setcctype{by}
\acmConference[SIGGRAPH Conference Papers '26]{Special Interest Group on Computer Graphics and Interactive Techniques Conference Conference Papers}{July 19--23, 2026}{Los Angeles, CA, USA}
\acmBooktitle{Special Interest Group on Computer Graphics and Interactive Techniques Conference Conference Papers (SIGGRAPH Conference Papers '26), July 19--23, 2026, Los Angeles, CA, USA}
\acmDOI{10.1145/3799902.3811054}
\acmISBN{979-8-4007-2554-8/2026/07}

\begin{document}

\title{EAG-PT: Emission-Aware Gaussians and Path Tracing for Diffuse Indoor Scene Reconstruction and Editing}

\author{Xijie Yang}
\orcid{0009-0009-3076-2595}
\affiliation{
    \institution{Zhejiang University}
    \city{Hangzhou}
    \country{China}}
\affiliation{
    \institution{Shanghai Artificial Intelligence Laboratory}
    \city{Shanghai}
    \country{China}}
\email{yangxijie@zju.edu.cn}

\author{Mulin Yu}
\orcid{0000-0002-0327-4547}
\authornote{Corresponding authors.}
\affiliation{
    \institution{Shanghai Artificial Intelligence Laboratory}
    \city{Shanghai}
    \country{China}}
\email{yumulin@pjlab.org.cn}

\author{Changjian Jiang}
\orcid{0000-0003-1774-1385}
\affiliation{
    \institution{Zhejiang University}
    \city{Hangzhou}
    \country{China}}
\email{changjianjiang01@gmail.com}

\author{Kerui Ren}
\orcid{0009-0003-8010-5733}
\affiliation{
    \institution{Shanghai Jiao Tong University}
    \city{Shanghai}
    \country{China}}
\affiliation{
    \institution{Shanghai Artificial Intelligence Laboratory}
    \city{Shanghai}
    \country{China}}
\email{renkerui@sjtu.edu.cn}

\author{Tao Lu}
\orcid{0009-0000-8830-3820}
\affiliation{
    \institution{Shanghai Artificial Intelligence Laboratory}
    \city{Shanghai}
    \country{China}}
\email{lutao@pjlab.org.cn}

\author{Jiangmiao Pang}
\orcid{0000-0002-6711-9319}
\affiliation{
    \institution{Shanghai Artificial Intelligence Laboratory}
    \city{Shanghai}
    \country{China}}
\email{pangjiangmiao@gmail.com}

\author{Dahua Lin}
\orcid{0000-0002-8865-7896}
\affiliation{
    \institution{The Chinese University of Hong Kong}
    \city{Hong Kong}
    \country{China}}
\affiliation{
    \institution{Shanghai Artificial Intelligence Laboratory}
    \city{Shanghai}
    \country{China}}
\email{dhlin@ie.cuhk.edu.hk}

\author{Bo Dai}
\orcid{0000-0003-0777-9232}
\authornotemark[1]
\affiliation{
    \institution{The University of Hong Kong}
    \city{Hong Kong}
    \country{China}}
\affiliation{
    \institution{Feeling AI}
    \city{Shanghai}
    \country{China}}
\email{bdai@hku.hk}

\author{Linning Xu}
\orcid{0000-0003-1026-2410}
\affiliation{
    \institution{The Chinese University of Hong Kong}
    \city{Hong Kong}
    \country{China}}
\affiliation{
    \institution{Shanghai Artificial Intelligence Laboratory}
    \city{Shanghai}
    \country{China}}
\email{linningxu@link.cuhk.edu.hk}

\renewcommand{\shortauthors}{Xijie Yang et al.}

\begin{abstract}
Recent radiance-field-based reconstruction methods, such as NeRF and 3DGS, achieve high visual fidelity for indoor scenes, but often break down under scene editing due to baked illumination and the lack of explicit light transport. In contrast, inverse path tracing methods based on mesh representations enforce correct light transport but require highly accurate geometry, making them difficult to apply robustly to real indoor scenes. We present Emission-Aware Gaussians and Path Tracing (EAG-PT), a method for physically based reconstruction and rendering of indoor scenes using a unified 2D Gaussian representation, targeting editable diffuse global illumination. Our approach consists of three key ideas: (1) representing indoor scenes with 2D Gaussians as a transport-friendly geometric proxy that avoids explicit mesh reconstruction; (2) explicitly separating emissive and non-emissive components during reconstruction to support editing; and (3) decoupling reconstruction from final rendering by using efficient single-bounce optimization and high-quality multi-bounce path tracing, respectively. Experiments on synthetic and real indoor scenes show that EAG-PT produces more natural and physically consistent edited renderings than radiance-field reconstructions, while preserving finer geometric detail and avoiding mesh-induced artifacts compared with mesh-based inverse path tracing. These results highlight the potential of our approach for applications such as interior design, XR content creation, and embodied AI.
\end{abstract}


\begin{teaserfigure}
    \vspace{-7pt} \large{\texttt{\href{https://eag-pt.github.io}{https://eag-pt.github.io}}} \vspace{5pt} \\
    \includegraphics[width=\textwidth]{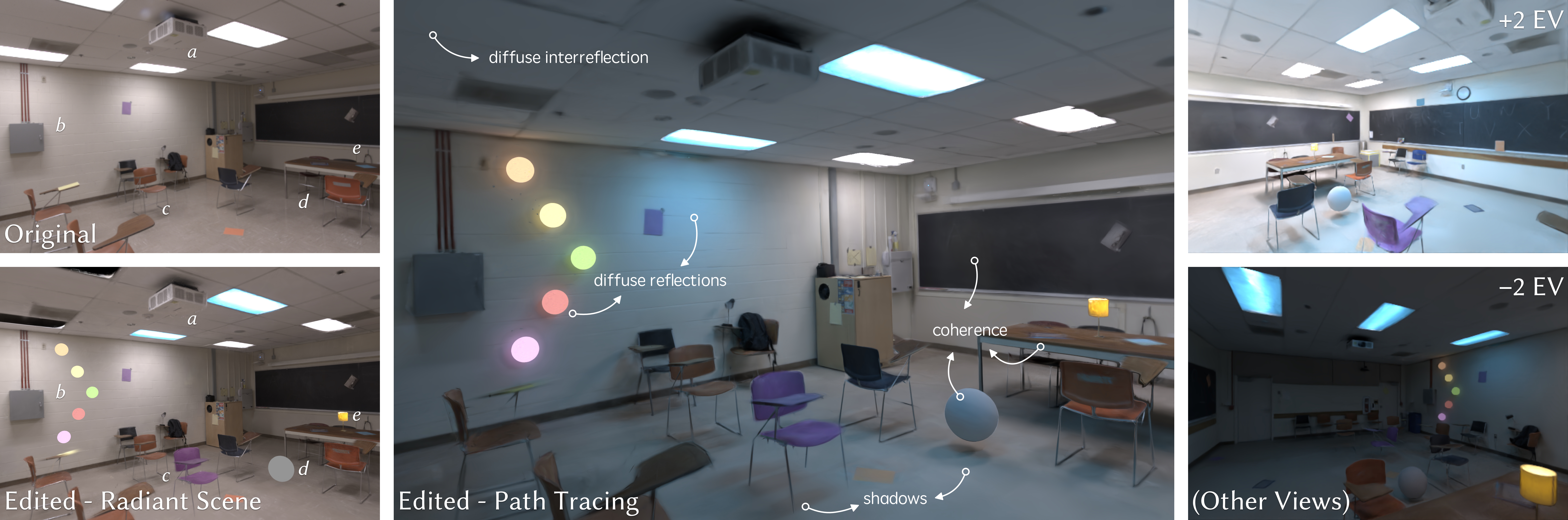}
    \caption{\label{fig:Teaser}
        Scene editing on 2D Gaussian primitives of a reconstructed real indoor scene, \textsc{f-classroom}, including:
        \textit{a}) relighting the ceiling,
        \textit{b}) inserting colorful emissive balls,
        \textit{c}) duplicating a chair with modified material,
        \textit{d}) adding a diffuse ball,
        and \textit{e}) importing a lamp from another scene.
        Path-traced rendering after editing produces coherent diffuse global illumination (reflections, interreflections, and shadows) in contrast to direct radiant scene composition.
    }
    \Description{Teaser}
\end{teaserfigure}

\maketitle

\section{INTRODUCTION}\label{sec:introduction}

Given multi-view captures of an indoor scene, modern 3D reconstruction methods such as Neural Radiance Fields (NeRF)~\cite{NeRF} and 3D Gaussian Splatting (3DGS)~\cite{3DGS} can recover scene representations that achieve high-fidelity novel-view synthesis. Compared to implicit neural representations and traditional mesh, the explicit Gaussian primitives used in 3DGS provide direct access to geometry and appearance parameters, making them attractive 3D representations for interactive scene manipulation and editing.
However, despite their representational flexibility, radiance-field-based reconstructions fail to produce physically consistent renders after scene editing. Modifying light sources, materials, or object layout does not yield corresponding changes in illumination or shadowing. This limitation stems from a shared modeling assumption: the whole scene is treated as uniformly radiant, with illumination implicitly encoded in outgoing radiance. While sufficient for reproducing the appearance at capture time, this formulation does not model light transport and therefore breaks under changes to scene configuration.

Prior efforts partially address this limitation by introducing limited reflection modeling~\cite{TexIR,I2SDF,EnvGS,EGR}. These methods add one or a small number of light bounces to improve view-dependent effects, yet they continue to rely on radiance cached from the original scene and do not explicitly reconstruct physical light sources. As a result, indirect illumination remains tied to the capture-time lighting configuration, and physically correct global illumination after editing remains out of reach.

At the other end of the spectrum, physically based inverse rendering~\cite{FIPT,IRIS} has long relied on mesh representations and path tracing to model light transport explicitly. While physically grounded, mesh-based inverse path tracing places strong requirements on geometric fidelity, which become a practical bottleneck for real indoor scenes with cluttered layouts and fine-scale structures. Errors in reconstructed meshes directly propagate through visibility, shading, and multi-bounce illumination, often dominating the final rendering quality. Recent work such as UGP~\cite{UGP} explores path tracing on Gaussian primitives, but focuses on forward rendering and does not address inverse reconstruction.

\begin{figure}
    \centering
    \includegraphics[width=\linewidth]{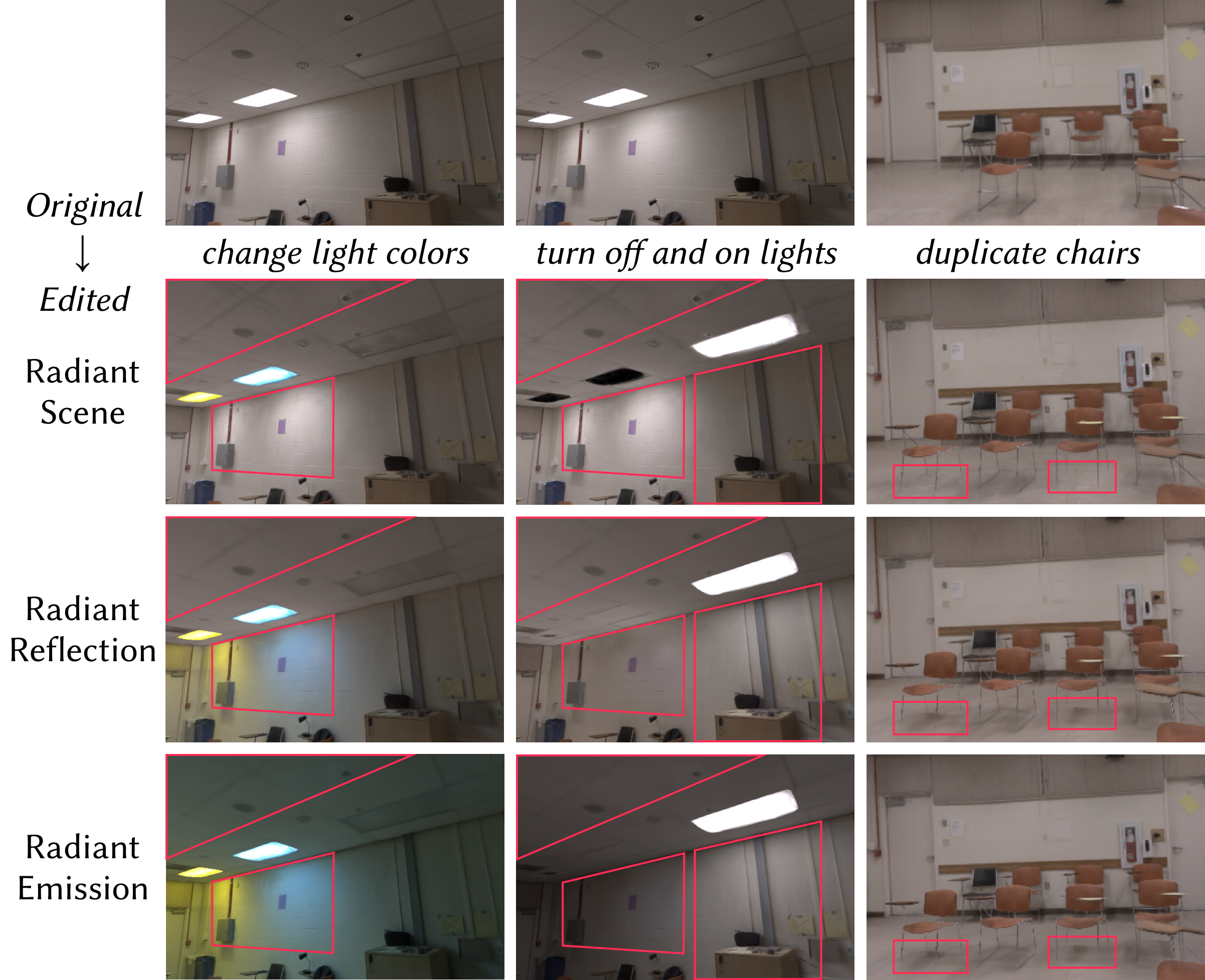}
    \caption{\label{fig:Introduction-issue-to-solve}
Renders of \textsc{f-classroom} before and after editing.
Radiant Scene: Most radiance field reconstruction works~\cite{NeRF,3DGS,3DGRT} regard the whole scene as radiant, which cannot produce light changes and shadow effects after scene editing.
Radiant Reflection: Some reflection modeling works~\cite{I2SDF,TexIR} add a single bounce to produce more realistic results, while still suffering from the incorrect radiance after scene editing.
Radiant Emission: We explicitly separate light sources from the radiant scene, and use path tracing to bounce light in the scene to derive photo-realistic renders. 
    }
    \Description{Introduction-issue-to-solve}
\end{figure}

To address these limitations, we propose Emission-Aware Gaussians and Path Tracing (EAG-PT), a physically grounded reconstruction framework that enables consistent indoor scene editing without mesh conversion. EAG-PT is built around four tightly coupled components, each targeting a core challenge in physically based scene reconstruction:

\begin{itemize}
\item \emph{Emission-aware scene decomposition}. We explicitly separate emissive light sources from non-emissive geometry using 2D emission masks, addressing the limitations of prior radiant-scene modeling for further editing.
\item \emph{Inverse recovery of radiance and material properties}. We recover emitter radiance and spatially varying diffuse surface reflectance for non-emissive components via differentiable rendering from multi-view observations.
\item \emph{Physically based light transport after editing}. We apply path tracing to edited scenes to re-evaluate multi-bounce diffuse global illumination, avoiding reliance on obsolete radiance cache at capture time.
\item \emph{Unified 2D Gaussian representation}. We adopt 2D Gaussians as a single scene representation that supports ray intersection, light transport, diffuse material modeling, and radiance caching, enabling efficient reconstruction and rendering.
\end{itemize}

Our method produces physically consistent renders with realistic diffuse global illumination after scene editing, as illustrated in Fig.~\ref{fig:Introduction-issue-to-solve}. Experiments on both synthetic and real scenes demonstrate clear improvements over prior radiance-field-based approaches for indoor scene reconstruction and editing. On real-world indoor scenes, comparisons with mesh-based inverse path tracing show that our unified Gaussian representation preserves finer geometric detail and yields more stable and visually natural results. Together, these results indicate that EAG-PT enables practical indoor scene editing, with broad applicability to interior design, XR content creation, and physically grounded asset preparation for embodied AI.

\section{RELATED WORK}\label{sec:related-work}

Below, we briefly review prior work on multi-view 3D reconstruction and inverse rendering for scenes, with a focus on indoor environments. We organize existing approaches by their treatment of light transport into three categories: radiant scene reconstruction, reflection modeling, and path tracing.

\subsection{Radiant Scene Reconstruction}\label{sec:related-work-radiant-scene-reconstruction}

Most multi-view 3D reconstruction methods implicitly treat the scene as radiant. Classical pipelines~\cite{SfM,MVS,Poisson,Metashape,Replica} reconstruct point clouds or meshes, attach them with view-independent colors, and render via rasterization. NeRF~\cite{NeRF} and its extensions~\cite{MipNeRF360,InstantNGP}, inspired by volume rendering~\cite{3DGS-volumetric-rendering-analysis}, instead represent the scene as a continuous radiance field optimized by ray marching and alpha blending, achieving significantly improved novel-view synthesis. 3DGS~\cite{3DGS} and its variants~\cite{2DGS,ScaffoldGS,OctreeGS,AnySplat} further model the scene as radiant 3D or 2D Gaussian primitives rendered by rasterization (splatting~\cite{EWA}) with alpha blending for high efficiency, while recent works~\cite{3DGRT,RadiantFoam,RaySplats,MeshSplats,RayGauss2025WACV,RayGaussX2025ICCV} adopt ray tracing on Gaussian primitives to alleviate some limitations of rasterization.

Despite differences in representation and rendering, these approaches share a key limitation: illumination is baked into appearance. As a result, they faithfully reproduce captured views but fundamentally cannot support physically plausible scene editing, such as modifying light sources or object layout. Our work departs from this formulation by explicitly separating emission from reflection and modeling multi-bounce light transport.

\subsection{Reflection Modeling}\label{sec:related-work-reflection-modeling}

Radiant methods model view-dependent effects using learned angular conditioning, such as MLP in NeRF~\cite{NeRF,SNISR,SpecNeRF} and spherical harmonics in 3DGS~\cite{3DGS,Plenoxels,InstantNGP}. While effective for specular effects, they still lack explicit light transport, limiting physically consistent scene editing.

Reflection-aware approaches separate scenes into a reflective base and an emitting environment, which works well for object-centric or local scenes~\cite{nvdiffrecmc,MIRReS,NeRFactor,R3DG,IRGS,PBIR-NeRF,EnvGS,MaterialRefGS,GaRe,FEGR}. For indoor environments, where emission and reflection are tightly coupled, recent methods jointly optimize materials and illumination within a single region~\cite{TexIR,I2SDF,EGR,IBL-NeRF,FVP}. However, these methods typically treat much of the scene as radiant and do not explicitly reconstruct light sources.

In contrast, we represent indoor scenes using 2D Gaussians~\cite{2DGS,EnvGS,MaterialRefGS,IRGS} and explicitly partition them into emissive and non-emissive components, enabling physically based light transport. The well-defined distance and normal of 2D Gaussians provide more stable geometry for light transport than volumetric 3D Gaussians~\cite{3DGS,3DGRT,EGR}.

\subsection{Path Tracing}\label{sec:related-work-path-tracing}

Path tracing~\cite{TheRenderingEquation} is the standard tool for simulating physically correct global illumination. Most prior indoor inverse rendering methods rely on meshes for geometry, either recovering materials and lighting directly in the mesh domain or parameterizing them with neural networks~\cite{InverseGI,IndoorEmptyingRefurnishingRelighting,FIPT-ref-IPT,FIPT-ref-33,Mitsuba3,FIPT-ref-MILO,FIPT,IRIS,IIF,FVP}. In these approaches, rendering quality is fundamentally constrained by mesh fidelity. For real-world indoor scenes, reconstructed meshes often fail to capture fine-scale geometry and are frequently converted from other representations such as SDFs or Gaussian primitives~\cite{2DGS,MonoSDF,GSDF,GOF}, introducing additional degradation.

By contrast, radiance field and Gaussian-based reconstructions offer more faithful geometric representations but are rarely integrated with physically based light transport.
Existing efforts include I2SDF~\cite{I2SDF} that adds a single bounce on NeRF and ESR-NeRF~\cite{ESR-NeRF} that considers emission modeling on objects, but they ignore path tracing for global illumination; UGP~\cite{UGP} adopts path tracing on Gaussian primitives for forward rendering yet does not address inverse reconstruction.

We aim to bridge this gap by enabling physically based light transport directly on Gaussian scene representations. We reconstruct emissive components and recover diffuse SVBRDF properties for non-emissive geometry, and apply multi-bounce path tracing after scene editing without mesh conversion. This formulation preserves fine-scale detail while maintaining physically consistent diffuse global illumination.

\section{METHOD}\label{sec:method}

\begin{figure*}
    \centering
    \includegraphics[width=\textwidth]{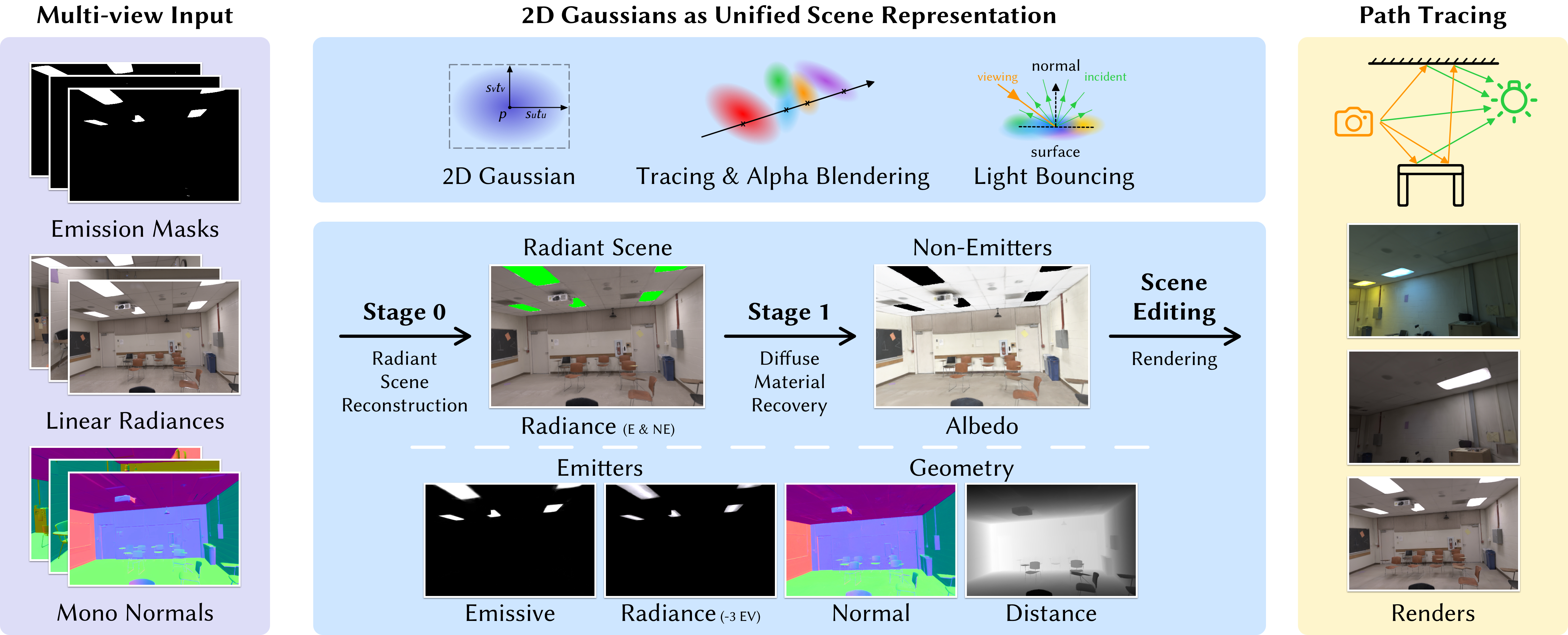}
    \caption{\label{fig:Method-pipeline}
Pipeline of Emission-Aware Gaussians and Path Tracing.
Given multi-view linear captures of an indoor scene with corresponding emitter masks and estimated normals, the radiant scene is first reconstructed in Stage~0 to get radiance, separate emitters, and derive geometry, based on 2D Gaussians and ray tracing.
The diffuse material of the non-emitters is then recovered in Stage~1 through light bouncing and differentiable rendering.
With properties of emitters, non-emitters, and scene geometry, path tracing that bounces light around the scene is adopted for photo-realistic renders on various scene editing scenarios.
    }
    \Description{Method-pipeline}
\end{figure*}

Our goal is to reconstruct a static indoor scene from multi-view images captured in linear color space with known camera poses, and to enable physically correct scene editing and photo-realistic path-traced rendering.
The input images densely cover the indoor environment, ensuring direct observation of light emitters.
As illustrated in Fig.~\ref{fig:Method-pipeline}, our pipeline proceeds in two stages. 
Stage~0, \emph{radiant scene reconstruction}, first lifts multi-view observations into a 3D representation of 2D Gaussians, followed by Stage~1, \emph{diffuse material recovery}, which estimates albedo for non-emissive regions.
After scene editing, path tracing is adopted on the derived scene for photo-realistic renders.

We begin in Sec.~\ref{sec:method-2d-gaussians} by introducing 2D Gaussians as the scene representation in our method, along with the associated tracing and bouncing used for rendering. We adopt 2D Gaussians as they provide a favorable trade-off between geometric accuracy and appearance quality.
In Sec.~\ref{sec:method-radiant-scene-reconstruction}, we describe radiant scene reconstruction, where the scene is initialized using differentiable rendering with tracing only, without any light bouncing.
Building on this initial radiance field, Sec.~\ref{sec:method-material-recovery} introduces the rendering equation and details diffuse material recovery for non-emissive regions via differentiable rendering with a single bounce into the radiance cache.
Finally, Sec.~\ref{sec:method-path-tracing} presents multi-bounce path tracing for physically based forward rendering, together with the corresponding light baking techniques.

\subsection{2D Gaussian Ray Tracing and Bouncing}\label{sec:method-2d-gaussians}

\paragraph{2D Gaussian Ray Tracing}

2DGS~\cite{2DGS} represents scenes using 2D Gaussians, which can be interpreted as small elliptical surface elements embedded in 3D space, enabling more geometrically compatible reconstruction over 3DGS. 
While subsequent works~\cite{IRGS,EnvGS} introduce limited ray tracing after splatting to query radiance from object surfaces or the environment, our method relies exclusively on ray tracing over 2D Gaussians. We therefore \emph{reformulate} 2D Gaussians directly as traceable primitives. Each 2D Gaussian is centered at a 3D position $\vec{p}$ and has finite spatial extent, with higher influence near its center and smoothly decaying weight toward its boundary. It is parameterized by anisotropic in-plane scale $s_{u,v}$, orientation represented by a quaternion $\hat{q}$, and opacity $\sigma$. Given a ray with origin $\vec{O}$ and direction $\hat{\omega}$, the intersection point in 3D space is given by $\vec{x} = \vec{O} + t\hat{\omega}$, where $t$ denotes the ray distance. The response of 2D Gaussian at the intersection is:
\begin{equation}
g(\vec{x}) = \exp \biggl\{ { - \frac{1}{2} \Bigl[ \bigl( \frac{(\vec{x}-\vec{p}) \cdot \hat{t}_u}{s_u} \bigr)^2 + \bigl( \frac{(\vec{x}-\vec{p}) \cdot \hat{t}_v}{s_v} \bigr)^2 \Bigr] }  \biggr\},
\end{equation}
\noindent
where $\hat{t}_{u,v}$ are unit vectors along the short and long axes derived from $\hat{q}$.

For a scene composed of 2D Gaussians, a ray keeps tracing forward and sequentially intersects with $n_\text{g}$ 2D Gaussians. At each intersection, a per-Gaussian micro-level quantity $v_i$ is accumulated via alpha blending to produce a macro-level quantity $V$, until the accumulated transparency $T$ falls below a preset threshold:
\begin{equation}
V = \sum_{i=1}^{n_\text{g}} w_i \cdot v_i
= \sum_{i=1}^{n_\text{g}} T_{i-1} \cdot \sigma_i g_i \cdot v_i, \ \ T_{i} = \prod_{j=1}^{i} (1 - \sigma_j g_j).
\end{equation}
In our formulation, the quantities include:
\begin{equation}
v \in \{ \pm \hat{n}, t, r, e, \rho \}, \ V \in \{ N, D, R, E, P \}.
\label{eq:2d-gaussians-ray-tracing-values}
\end{equation}
\noindent For each 2D Gaussian (and a given ray), $\pm \hat{n} = \pm \hat{t}_u \times \hat{t}_v$ ($N$) denotes the surface normal oriented toward the camera center, $t$ ($D$) is the intersection distance, $r$ ($R$) is the linear radiance, $e$ ($E$) is the emissive term in $[0,1]$, and $\rho$ ($P$) is the diffuse albedo.
The accumulated normal and distance are normalized as $\tilde{N} = {N} / {||N||},\ \tilde{D} = {D} / {A}$, where $A = 1 - T_n$.

\paragraph{Light Bouncing}

After tracing, to bounce the ray, the effective macro-level intersection point is defined as $\vec{X} = \vec{O} + \tilde{D} \hat\omega$ for the new origin. And the new direction $\hat\omega'$ is sampled from the upper hemisphere $\Omega^+$ defined by $\tilde{N}$, with sampling probability $p(\hat\omega')$.
The new ray then proceeds and collects a new set of tracing results $V' \in \{ \tilde{N}', \tilde{D}', R', E', P' \}$, and $\vec{X}' = \vec{X} + \tilde{D}'\hat\omega'$, 
after which the process either terminates or starts another bounce.
Note that, light bounces are not applied when reconstructing radiant scene~(Sec.~\ref{sec:method-radiant-scene-reconstruction}), while a single bounce is added for diffuse material recovery~(Sec.~\ref{sec:method-material-recovery}) and multiple bounces are required for path tracing~(Sec.~\ref{sec:method-path-tracing}).
For simplicity, we omit the intersection point and only use direction $\hat\omega$ to represent a ray in subsequent equations.

\subsection{Radiant Scene Reconstruction}\label{sec:method-radiant-scene-reconstruction}

Given multi-view inputs and an initial coarse RGB point cloud of an indoor scene, we first lift 2D observations into a 3D radiant scene represented by a collection of 2D Gaussians.

\paragraph{Radiance Loss}

At Stage~0, the scene is radiant, which means outgoing radiance $L_o^0(\hat\omega_o)$ of a pixel is radiance $R(\hat\omega_o)$ of the ray.
Following prior work on radiant scene reconstruction, we employ the standard color reconstruction loss $\mathcal{L}_c$~\cite{3DGS,2DGS,3DGRT} for radiance.
To improve numerical stability when optimizing linear radiance values, we apply a perceptual quantization curve $(\cdot)_\text{PQ}$~\cite{SMPTE-ST-2084,VR-NeRF}, resulting in $\mathcal{L}_c^0 = \mathcal{L}_c \bigl( ( L_o^0 )_\text{PQ}, ( L^{\text{gt}}_{o} )_\text{PQ} )$.

\paragraph{Geometry Loss}

By only applying radiance loss, the reconstructed radiant scene looks photo-realistic, but generally with poor geometry~\cite{EGR,ScaffoldGS}, which is insufficient for precise light bounce.
To improve geometric fidelity, we follow~\cite{2DGS,EnvGS,IRIS} and supervise both surface orientation and depth variation. Specifically, render normal $\tilde{N}$ and distance normal $\tilde{N}_d$ (the gradient of $\tilde{D}$) are directly supervised by mono normal maps estimated from sRGB images using StableNormal~\cite{StableNormal}: $\mathcal{L}_{n} = || 1- \tilde{N} \cdot N_{\text{mono}} ||_1, \ \mathcal{L}_{d} = || 1- \tilde{N}_d \cdot N_{\text{mono}} ||_1$.
Empirically, we observe that even state-of-the-art monocular depth estimators~\cite{DepthAnythingv2,MoGev2} lack the accuracy and cross-view consistency required for reliable supervision, whereas monocular normal estimation provides more stable and geometrically meaningful guidance.

\paragraph{Emission Loss}

To explicitly distinguish physical light sources from reflective surfaces, we incorporate 2D emission masks for scene editing and path tracing.
Given an emission mask $M$, we supervise the emissive component $E$ via $\mathcal{L}_e = || E - M ||_1$.
Details on the construction of 2D emission masks are provided in Appendix~\ref{sec:appendix-emission-masks}.

\paragraph{Final Loss}

We jointly optimize the parameters of all 2D Gaussians using differentiable 2D Gaussian ray tracing by minimizing the weighted sum of the above losses given weights $\lambda_{c},\lambda_{n},\lambda_{d},\lambda_{e}$:

\begin{equation}
\min_{\vec{p},s,\hat{q},\sigma,r,e}\mathcal{L}^0 = \lambda_{c} \mathcal{L}_c^0 + \lambda_{n} \mathcal{L}_{n} + \lambda_{d} \mathcal{L}_{d} + \lambda_{e} \mathcal{L}_{e}.
\label{eq:optimization-stage-0}
\end{equation}

After this stage, a radiant scene is reconstructed, with accurate geometry for light bouncing, separation of emitters and non-emitters, true radiance of emitters, and radiance cache of non-emitters. 
This representation serves as the foundation for diffuse material recovery with a single bounce~(Sec.~\ref{sec:method-material-recovery}) and multi-bounce path tracing~(Sec.~\ref{sec:method-path-tracing}).

\subsection{Diffuse Material Recovery via Single Bounce}\label{sec:method-material-recovery}

With the reconstructed radiant scene obtained in Sec.~\ref{sec:method-radiant-scene-reconstruction}, we perform single-bounce differentiable rendering to recover diffuse material properties of the 2D Gaussians (Stage~1 in Fig.~\ref{fig:Method-pipeline}).
We first review the rendering equation under our assumptions, and then describe the material recovery procedure enabled by a single bounce with radiance cache.

\paragraph{The Rendering Equation}

The rendering equation~\cite{TheRenderingEquation} depicts light transport inside 3D space at each surface point.
Since emission typically dominates reflection, we follow~\cite{FIPT,IRIS} and assume that emissive surfaces do not reflect incoming light, enabling a clean separation between emission and non-emissive reflection:

\begin{equation}
\begin{aligned}
L_o(\hat\omega_o)
&= L_e(\hat\omega_o) + L_r(\hat\omega_o) \\
&= L_e(\hat\omega_o) \ \text{ if } \ E(\hat\omega_o) > \tau_E \ \text{ else } \ L_r(\hat\omega_o),
\end{aligned}
\label{eq:the-rendering-equation-assumption}
\end{equation}

\noindent
where $\hat\omega_o$ is the viewing direction, $L_o$ the outgoing radiance, $L_e$ the radiance from emitters, and $L_r$ the reflected radiance. $\tau_E=0.1$ is used to keep a smooth transition between emitters and non-emitters, larger values creating prominent boundaries and smaller ones introducing redundant emission. For emitters, $L_e(\hat\omega_o) = R(\hat\omega_o)$. For reflection of non-emitters:

\begin{equation}
L_r(\hat\omega_o)
= \int_{\Omega^+}
  L_i(\hat\omega_i) \ f(\hat\omega_i, \hat\omega_o) \ (\hat\omega_i \cdot \tilde{N})
  \ \mathrm{d} \hat\omega_i,
\label{eq:the-rendering-equation-reflection}
\end{equation}

\noindent
where $\hat{\omega}_i$ is the incident direction, $L_i$ the incident radiance, and $f$ the BRDF. This formulation is recursive, as $L_i$ itself depends on radiance reflected from other surfaces.

\paragraph{Diffuse Material Recovery}

For material recovery, we follow the idea of using radiance cache (or irradiance cache) from \cite{DiffuseInterreflection,NeuralRadianceCache,I2SDF,TexIR}.
By approximating the incident radiance $L_i (\hat\omega_i)$ using the accumulated radiance $R (\hat\omega_i)$ obtained from the radiant scene reconstruction stage, we remove the need for iterative multi-bounce simulation during material recovery.
The reflected radiance is estimated via Monte Carlo integration with $n_\text{spp}$ samples per pixel:

\begin{equation}
L_r^1(\hat\omega_o) \approx
  \frac{1}{n_\text{spp}} \sum^{n_\text{spp}}
  R(\hat\omega_i)
  \frac{f(\hat\omega_i, \hat\omega_o) \ (\hat\omega_i \cdot \tilde{N})}{p(\hat\omega_i)} 
  , \ \hat\omega_i \sim \Omega^+.
\end{equation}
In this work, we adopt only the diffuse BRDF $f(\hat\omega_i, \hat\omega_o) = P / \pi$. The extension beyond diffuse material is discussed in Appendix~\ref{sec:appendix-discussion}. Diffuse albedo $\rho$ of each 2D Gaussian is optimized by:
\begin{equation}
\min_\rho \mathcal{L}^1 = \lambda_c \mathcal{L}_c \bigl( ( L_o^1 )_\text{PQ}, ( L_o^\text{gt} )_\text{PQ} \bigr).
\label{eq:optimization-stage-1}
\end{equation}

\noindent
Notice that radiance $r$ is only optimized in Stage~0~(Sec.~\ref{sec:method-radiant-scene-reconstruction}) and kept fixed in this stage to avoid the diffuse-emission ambiguity, as pointed out in~\cite{FIPT,FIPT-ref-MILO}.

\subsection{Path Tracing for Scene Editing}\label{sec:method-path-tracing}

\paragraph{Path Tracing}

After scene editing, previous works \cite{3DGRT,I2SDF,EGR,TexIR} that do not use path tracing, still use $L_o^0(\hat\omega_o)$ or $L_o^1(\hat\omega_o)$ with obsolete radiance cache $R$ at capture time for final rendering.
However, scene editing (e.g. changing light colors, inserting objects, etc.) should change $R$, and usage of obsolete radiance cache produces unnatural rendering results.

In our method, we only \emph{keep} accurate radiance of emitters, \emph{drop} obsolete radiance cache of non-emitters, and \emph{adopt path tracing} to derive final renders after scene editing.
This should correctly solve the recursion in Eqs.~\ref{eq:the-rendering-equation-assumption},\ref{eq:the-rendering-equation-reflection}.
Each ray shoots from the camera center, intersects with 2D Gaussians at $\vec{X_1}$, and bounces around inside the scene at $\vec{X}_2, \cdots, \vec{X}_b$, until $\vec{X}_b$ is emissive. The ray is discarded if bounce count exceeds the given bounce limit $\tau_b$, and valid paths are averaged to reduce sampling noise:

\begin{equation}
\begin{gathered}
L_r^\text{pt}(\hat\omega_o) \approx
  \frac{1}{n_\text{spp}} \sum^{n_\text{spp}}
  L_e(\hat\omega_{i,b})
  \prod^b_{k=1}\frac{f(\hat\omega_{i,k}, \hat\omega_{o,k}) \ (\hat\omega_{i,k} \cdot \bar{N}_k)}{p(\hat\omega_{i,k})} 
  , \ \hat\omega_{i,k} \sim \Omega^+_k, \\
L_e(\hat\omega_{i,b}) = R(\hat\omega_{i,b}) \ \text{ if } \ \bigl( E(\hat\omega_{i,b}) > \tau_E \ \text{ and } \ b \le \tau_b \bigr) \ \text{ else } \ 0.
\end{gathered}
\end{equation}

\noindent
$L_o^\text{pt}$ is finally sent to a denoiser~\cite{OptiXDenoiser} for better visual quality.

\paragraph{Light Baking}

While path tracing produces physically accurate renderings after scene editing, its computational cost precludes real-time performance, and low sample counts often result in visible noise or blur.
To enable efficient visualization of edited scenes, we adopt a light baking strategy inspired by commercial game engines~\cite{LightBakingSystem}.
Specifically, we re-bake the radiance obtained from path tracing into the 2D Gaussian representation by directly optimizing the per-Gaussian radiance $r$:
\begin{equation}
\min_{r}\mathcal{L}^\text{lb} = \lambda_c \mathcal{L}_c \bigl(  ( L_o^0 )_\text{PQ}, ( L_o^\text{pt} )_\text{PQ} \bigr).
\label{eq:optimization-light-baking}
\end{equation}
\noindent
This transfers global illumination effects into the radiant scene representation, enabling interactive real-time 0-bounce rendering of edited scenes without a denoiser and, in practice, slightly reducing residual blur.

\section{EXPERIMENTS}\label{sec:experiments}

\begin{table*}[b]
    \caption{\label{tab:before-and-after-scene-editing}
        Results on synthetic scenes before and after relighting.
        Path tracing achieves the highest relighting quality on synthetic scenes, and light baking preserves this quality while greatly reducing render time.
    }
\scalebox{0.875}{
    \begin{tabular}{ l | ccc | cccc | ccc | cccc }
    \toprule
    
    Scene             & \multicolumn{7}{c|}{\textsc{b-kitchen}}
                      & \multicolumn{7}{c}{\textsc{b-livingroom}} \\
    Setting           & \multicolumn{3}{c|}{\color{gray} \textit{original}} & \multicolumn{4}{c|}{\textit{relighted}}
                      & \multicolumn{3}{c|}{\color{gray} \textit{original}} & \multicolumn{4}{c}{\textit{relighted}} \\
    Method            & \color{gray} $\text{PSNR}^\uparrow$ & \color{gray} $\text{LPIPS}^\downarrow$ & \color{gray} $\text{FLIP}^\downarrow$
                      & $\text{PSNR}^\uparrow$ & $\text{LPIPS}^\downarrow$ & $\text{FLIP}^\downarrow$ & $\text{Time}^\downarrow$
                      & \color{gray} $\text{PSNR}^\uparrow$ & \color{gray} $\text{LPIPS}^\downarrow$ & \color{gray} $\text{FLIP}^\downarrow$
                      & $\text{PSNR}^\uparrow$ & $\text{LPIPS}^\downarrow$ & $\text{FLIP}^\downarrow$ & $\text{Time}^\downarrow$ \\
    
    \hline
    
    0-Bounce          & \color{gray} \textbf{37.57} & \color{gray} \textbf{0.0680} & \color{gray} \textbf{0.0732}
                      & 16.22 & 0.1098 & 0.3593 & \textbf{0.015}
                      & \color{gray} \textbf{37.83} & \color{gray} \textbf{0.0587} & \color{gray} \textbf{0.0646}
                      & 11.59 & 0.1687 & 0.5147 & \textbf{0.013} \\
    1-Bounce          & \color{gray} 32.05 & \color{gray} 0.0733 & \color{gray} 0.1064
                      & 19.78 & 0.0896 & 0.3176 & 27.9
                      & \color{gray} 33.02 & \color{gray} 0.0724 & \color{gray} 0.0982
                      & 15.47 & 0.1243 & 0.4474 & 22.4 \\

    \hline
                      
    Ours (Path Tracing) & \color{gray} 26.57 & \color{gray} 0.0829 & \color{gray} 0.1759    
                        & 28.70 & \textbf{0.0825} & 0.1839 & 188
                        & \color{gray} 27.03 & \color{gray} 0.0829 & \color{gray} 0.1843
                        & 29.30 & \textbf{0.0983} & 0.2047 & 155 \\

    Ours (Light Baking) & \color{gray} - & \color{gray} - & \color{gray} -
                        & \textbf{28.97} & 0.0977 & \textbf{0.1823} & \textbf{0.015}
                        & \color{gray} - & \color{gray} - & \color{gray} -
                        & \textbf{29.33} & 0.1033 & \textbf{0.2045} & \textbf{0.013} \\
    
    \bottomrule
    \end{tabular}
}
\end{table*}

\begin{figure*}[b]
    \centering
    \includegraphics[width=\textwidth]{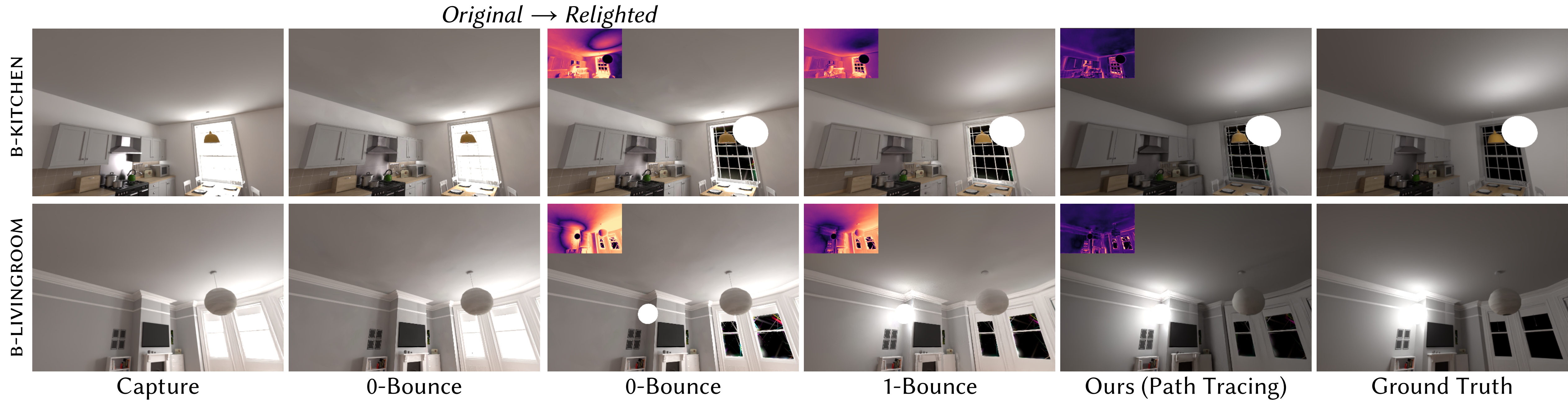}
    \caption{\label{fig:Result-synthetic-lightball-relighting}
Relighting results with an inserted illuminated ball on synthetic scenes. 
Insets show FLIP error maps w.r.t. the relighting ground truth.
While 0-bounce and 1-bounce renderings fail to reproduce global illumination after editing, our path tracing reproduces the target diffuse global illumination.
    }
    \Description{Result-synthetic-lightball-relighting}
\end{figure*}

\begin{figure*}
    \centering
    \includegraphics[width=\textwidth]{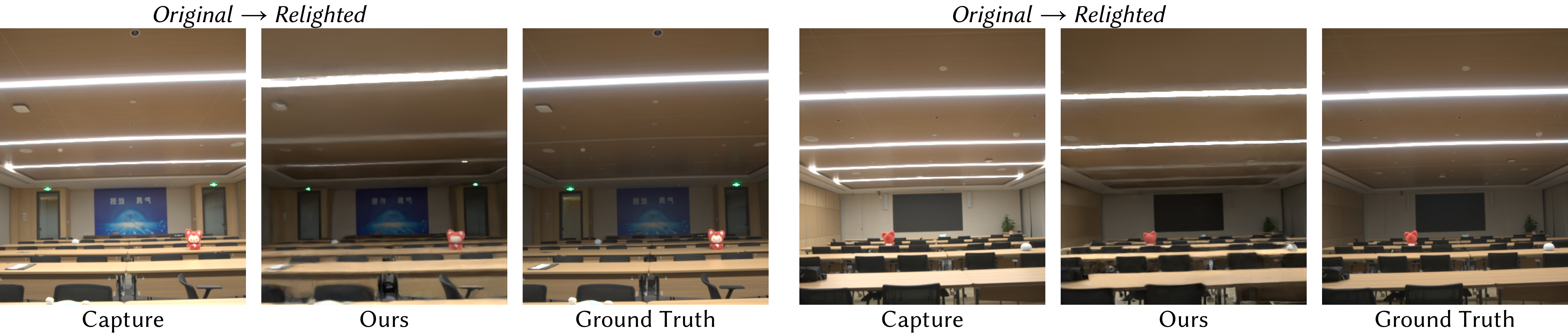}
    \caption{\label{fig:Result-real-lectureroom-relighting}
Relighting results on the captured real scene \textsc{lectureroom}.
For each relighting condition that turns off half lights, our path tracing closely reproduces the spatially varying indoor illumination compared with ground-truth relit photograph.
    }
    \Description{Result-real-lectureroom-relighting}
\end{figure*}

\begin{figure*}
    \centering
    \includegraphics[width=\textwidth]{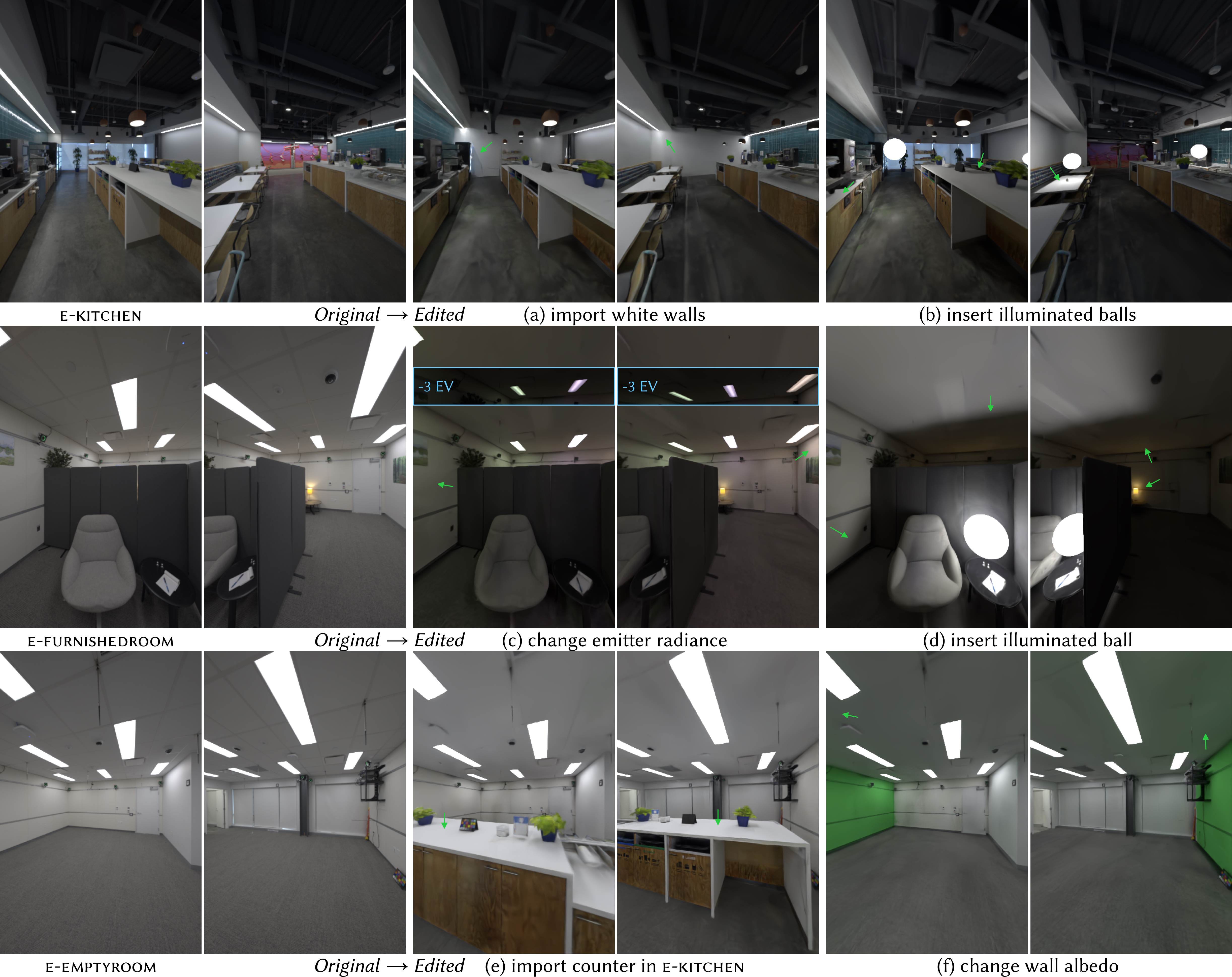}
    \caption{\label{fig:Result-real-eyefultower-editing.jpg}
Path-traced results for various scene-editing operations on Eyeful Tower scenes. After editing, our EAG-PT yields plausible renders: imported non-emissive objects integrate naturally into the scene (a,e), inserted luminous balls cast consistent reflections and shadows (b,d), and modified emitters or materials produce the expected changes in atmosphere (c,f). Continuous renderings and comparisons to the counterintuitive 0-bounce (radiant scene composition) baseline are provided in the supplementary video.
    }
    \Description{Result-real-eyefultower-editing}
\end{figure*}

\subsection{Datasets}

Since path tracing is performed in linear radiance space, our method requires multi-view indoor captures with calibrated linear radiance. For real scenes, this is typically obtained via exposure bracketing followed by HDR merging.
We primarily evaluate our method on real-world datasets. Specifically, we use indoor scenes from FIPT~\cite{FIPT} (\textsc{f-}) and VR-NeRF Eyeful Tower~\cite{VR-NeRF} (\textsc{e-}).
For \textsc{f-}, each scene contains several hundred views at a resolution of $360 \times 540$, with $1/8$ of the views reserved for testing.
For \textsc{e-}, each scene includes thousands of views captured by a camera rig, downsampled to $540 \times 360$.
For completeness, we also include two synthetic scenes from~\cite{BB-resources}, directly exported from Blender (\textsc{b-}) with ground-truth relighting results obtained by inserting a light ball.
As existing real-world datasets do not provide relighting ground truths, we capture an additional scene, \textsc{lectureroom}, with controlled relighting for comprehensive validation. Further details on data acquisition are provided in Appendix~\ref{sec:appendix-capture}.

\subsection{Implementation Details}

Drawing inspiration from~\cite{3DGRT,EnvGS,IRGS,TorchOptiX}, we implement differentiable 2D Gaussian ray tracing with ray bouncing, including Stage~0, Stage~1, and path tracing, from scratch using PyTorch and OptiX~\cite{OptiX}. This hybrid implementation is significantly faster than a PyTorch-only baseline in practice.
We fix the number of 2D Gaussians to reflect scene complexity: 200k for \textsc{b-}, 500k for \textsc{f-} and \textsc{lectureroom}, and 1M for \textsc{e-}.
In Stage~0, we reconstruct the radiant scene using 30k iterations with $\lambda_c = 1.0$, $\lambda_n = 0.5$, $\lambda_d = 0.05$, and $\lambda_e = 0.1$.
In Stage~1, diffuse albedo is optimized for 400 iterations with $n_\text{spp}=256$.
For rendering, we use $n_\text{spp}=1024$ for both single-bounce (1-bounce) and path tracing to reduce noise, with a bounce limit of $\tau_b = 7$ for path tracing.
During light baking, path-traced renders on all train set views are used to optimize per-Gaussian radiance for 3k iterations.
All experiments are conducted on a single NVIDIA RTX~4090 GPU.
The code is open-source and available at \mbox{\url{https://github.com/InternRobotics/EAG-PT}} to support reproducibility and further research.

\subsection{Results}

\subsubsection{Results on Synthetic Scenes}

We report quantitative comparisons of different rendering strategies before (\textit{original}) and after (\textit{relighted}) inserting an illuminated ball on the synthetic scenes \textsc{b-kitchen} and \textsc{b-livingroom} in Table~\ref{tab:before-and-after-scene-editing}, for which relighting ground truth is available.
We evaluate FLIP~\cite{FLIP} on linear-radiance images, and PSNR and LPIPS~\cite{LPIPS} on sRGB images obtained by converting linear radiance to sRGB and clipping to $[0,1]$. Qualitative comparisons are shown in Fig.~\ref{fig:Result-synthetic-lightball-relighting}.

On the original scenes, radiant scene reconstruction without light bounces (0-bounce) achieves the best performance, as it directly recovers the radiance at capture time. Introducing additional bounces (1-bounce or path tracing) degrades reconstruction accuracy.
After scene editing, 0-bounce rendering fails to model physically correct light transport and produces visually inconsistent results. Incorporating a single bounce provides limited improvement but remains insufficient to achieve diffuse global illumination. Only emission-aware multi-bounce path tracing consistently yields photo-realistic relighting results.

While path tracing produces the highest visual fidelity, it incurs rendering costs on the order of hundreds of seconds per frame. To address this limitation for content distribution and interactive applications on edited scenes, we re-bake the radiance in path-traced renders into 2D Gaussians and render using 0-bounce. As shown in Table~\ref{tab:before-and-after-scene-editing}, this light-baking strategy uses only one-tenth as many iterations as Stage~0, while preserving rendering quality and dramatically improving rendering speed over path tracing, thereby enabling real-time navigation in edited scenes.

\subsubsection{Results on Real Scenes}

\begin{figure*}[b]
    \centering
    \includegraphics[width=\textwidth]{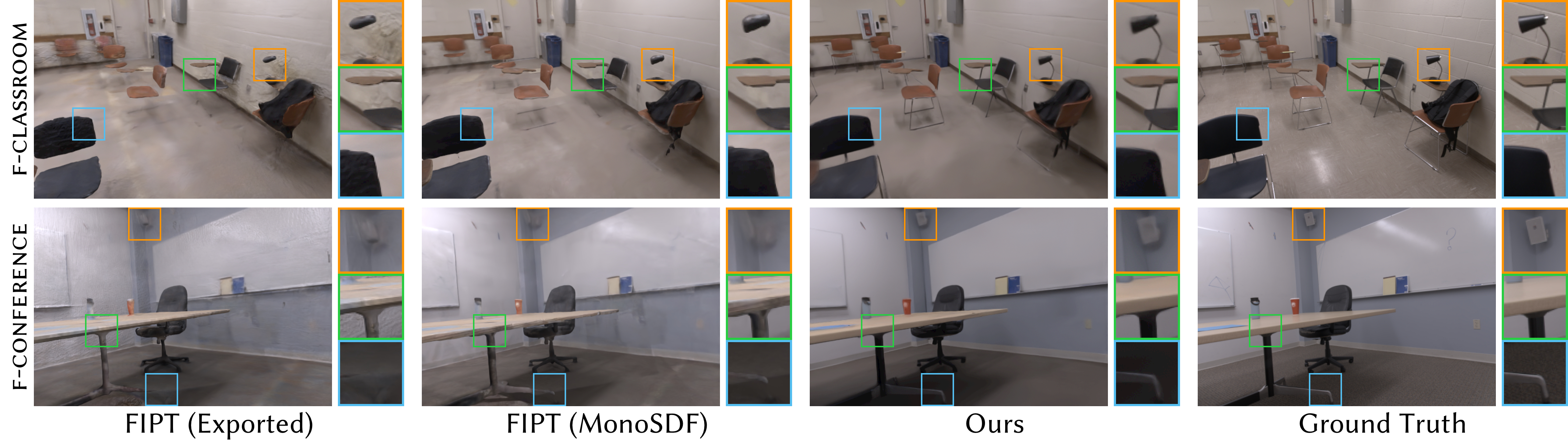}
    \caption{\label{fig:Result-real-comparison-with-fipt}
Path-traced novel views on real scenes compared with mesh-based FIPT. Zoomed regions highlight that our 2D Gaussians better preserve thin structures and avoid mesh triangulation artifacts, yielding more detailed and natural renderings.
    }
    \Description{Result-real-comparison-with-fipt}
\end{figure*}

\begin{table}[b]
    \caption{\label{tab:comparison-with-fipt}
        Our Gaussian-based path tracing achieves consistently better novel-view quality than mesh-based FIPT on real scenes.
    }
\scalebox{0.80}{
    \begin{tabular}{ l | ccc | ccc }
    \toprule
    
    Scene             & \multicolumn{3}{c|}{\textsc{f-classroom}}
                      & \multicolumn{3}{c}{\textsc{f-conference}} \\
    Method            & $\text{PSNR}^\uparrow$ & $\text{LPIPS}^\downarrow$ & $\text{FLIP}^\downarrow$
                      & $\text{PSNR}^\uparrow$ & $\text{LPIPS}^\downarrow$ & $\text{FLIP}^\downarrow$ \\

    \hline
    
    \color{gray} 0-Bounce             & \color{gray} 32.09 & \color{gray} 0.1432 & \color{gray} 0.1329 
                         & \color{gray} 29.41 & \color{gray} 0.1809 & \color{gray} 0.1326 \\
    \color{gray} 1-Bounce             & \color{gray} 30.37 & \color{gray} 0.1849 & \color{gray} 0.1658
                         & \color{gray} 28.57 & \color{gray} 0.1834 & \color{gray} 0.1490 \\
    
    \hline

    FIPT (Exported)      & 23.57 & 0.3042 & 0.2801
                         & 21.54 & 0.3948 & 0.2856 \\
    FIPT (MonoSDF)       & 26.38 & 0.2031 & 0.2265
                         & 22.29 & 0.2729 & 0.2622 \\
    Ours (Path Tracing)         & \textbf{28.65} & \textbf{0.1998} & \textbf{0.2117}
                         & \textbf{26.44} & \textbf{0.1960} & \textbf{0.2066} \\

    \bottomrule
    \end{tabular}
}
\end{table}

Path tracing results on the captured real-world scene \textsc{lectureroom}, compared against relighting ground truth, are shown in Fig.~\ref{fig:Result-real-lectureroom-relighting}.
Our method faithfully reconstructs the scene under the fully illuminated condition and produces visually consistent relighting results when half of the light sources are turned off.
We further demonstrate a range of scene editing operations on real scenes, including \textsc{f-classroom}~(Figs.~\ref{fig:Teaser},\ref{fig:Introduction-issue-to-solve}) and Eyeful Tower scenes~(Fig.~\ref{fig:Result-real-eyefultower-editing.jpg}). These edits include modifying lighting and material properties, inserting illuminated balls, and importing non-emissive objects.
Across all scenarios, path tracing produces coherent renderings with natural reflections, physically plausible shadows, interreflections, and consistent diffuse global illumination.
We encourage readers to refer to the supplementary video for direct visual comparisons between counterintuitive 0-bounce rendering (radiant scene composition) and physically plausible path tracing. In addition to static scene edits, the supplementary video also presents dynamic editing results, including moving illuminated balls in \textsc{f-} scenes.

\subsubsection{Comparison with FIPT}

We compare our method with the state-of-the-art mesh-based inverse path tracing approach FIPT~\cite{FIPT} on the real scenes \textsc{f-classroom} and \textsc{f-conference}.
FIPT operates on reconstructed meshes; we evaluate two mesh variants: (i) the MonoSDF mesh~\cite{MonoSDF} provided by the FIPT authors, trained for approximately one day per scene, and (ii) a TSDF mesh (resolution 512) reconstructed from multi-view depth maps derived from our representation, which exhibits comparable geometry while requiring only several minutes to generate.
The optimization times of the two methods are comparable on an RTX~4090 (ours Stage~0 + Stage~1: 19+32=51~min and 17+33=50~min; FIPT: 86~min and 61~min).
Quantitative novel-view path tracing results with $n_\text{spp}=1024$ are reported in Table~\ref{tab:comparison-with-fipt}, with qualitative comparisons shown in Fig.~\ref{fig:Result-real-comparison-with-fipt}.

Our method consistently achieves higher rendering quality than FIPT. Mesh representations struggle to capture fine geometric details such as chair legs and lamp arms in real-world scenes, even at high resolution. In contrast, 2D Gaussians naturally model such structures using anisotropic primitives. Moreover, mesh-based path tracing often exhibits visible triangulation artifacts at edges and corners, whereas our Gaussian-based representation produces smoother and more realistic results.
Additionally, our method adopts 2D Gaussians as the unified representation, which is much simpler than FIPT that combines triangle mesh, voxel grid, MLP material, and image-based shading.
Besides the mesh that occupies around 150 MB storage per scene, FIPT also uses around 500 MB to store recovered material (and additional 23 GB storage for image-based shading during training). While our method produces only a single 33 MB ply file that stores 500k 2D Gaussians containing all properties, which is only 5\% of compared to FIPT.
GPU memory also remains below 2.5 GB throughout, further highlighting the efficiency of our method.

\subsection{Ablations}

\begin{table}[b]
    \caption{\label{tab:ablations}
        Ablation study on \textsc{f-classroom} comparing rendering quality and speed across different configurations.
    }
\scalebox{0.97}{
    \begin{tabular}{ l | cccc }
    \toprule
    
    Method            & $\text{PSNR}^\uparrow$ & $\text{LPIPS}^\downarrow$ & $\text{FLIP}^\downarrow$ & $\text{Time}^\downarrow$ \\

    \hline

    w/o normal supervision      & 28.29 & 0.2122 & 0.2158 & 160  \\
    w/o normal consistency      & 29.20 & 0.1952 & 0.1978 & 125  \\

    \hline

    inaccurate emission mask    & 23.58 & 0.2162 & 0.3232 & 132 \\
    
    \hline
    
    bounce limit $\tau_b$ 7 $\rightarrow$ 3       & 24.94 & 0.2092 & 0.3277 & 55  \\
    bounce limit $\tau_b$ 7 $\rightarrow$ 11      & 28.90 & 0.1989 & 0.2029 & 187 \\
    
    \hline
    
    $n_\text{spp}$ 1024 $\rightarrow$ 256           & 28.57 & 0.2283 & 0.2131 & 31  \\
    $n_\text{spp}$ 1024 $\rightarrow$ 4096          & 28.67 & 0.1846 & 0.2113 & 496 \\

    \hline

    lower Gaussian count                 & 28.39 & 0.2187 & 0.2152 & 110 \\
    
    \hline

    Full  & 28.65 & 0.1998 & 0.2117 & 123 \\
    
    \bottomrule
    \end{tabular}

}
\end{table}

\begin{figure*}[b]
    \centering
    \includegraphics[width=\textwidth]{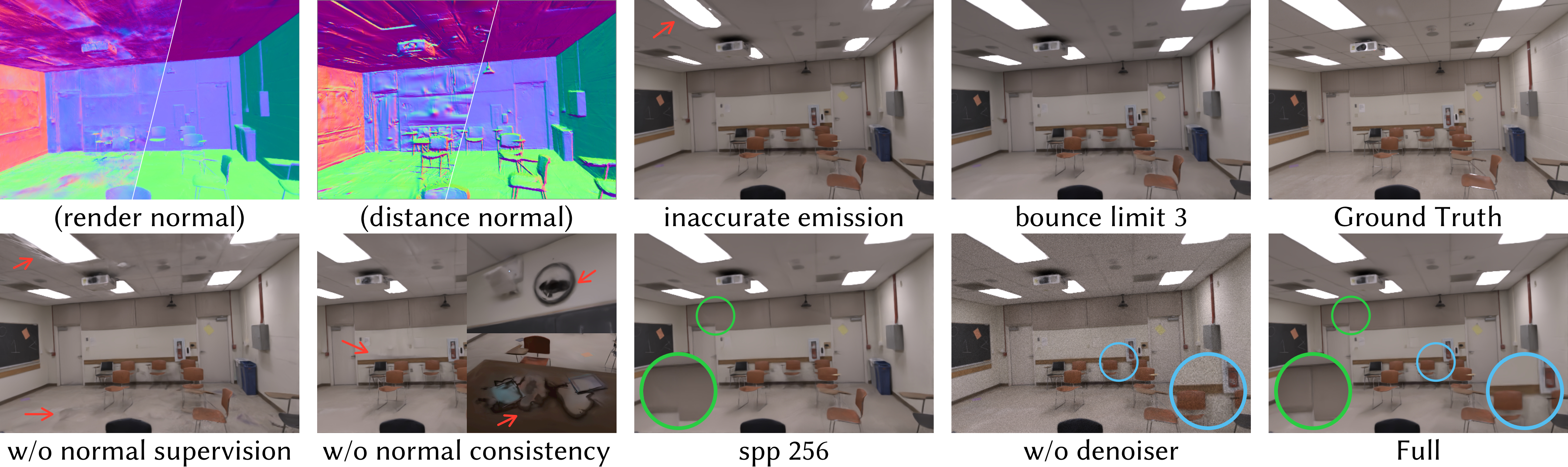}
    \caption{\label{fig:Ablation}
Ablation study on \textsc{f-classroom}. The comparisons show that accurate normals, proper emission masks, sufficient bounce limit and samples per pixel, and a denoiser are all necessary to avoid artifacts and to achieve our final high-quality path-traced results.
    }
    \Description{Ablation}
\end{figure*}

\paragraph{Normal Supervision and Consistency}

Correct light bouncing in our method depends on good geometry (per-pixel normal and distance). As shown in Fig.~\ref{fig:Ablation}, normal supervision and normal consistency together produce smooth surfaces with the help of estimated normal maps.
The normal supervision prevents extruding Gaussians, and normal consistency avoids cavities in the scene (though better quantitative numbers are achieved without applying normal consistency in Table~\ref{tab:ablations}).

\paragraph{Emission Masks}

Emission masks play a significant role in separating emitters and non-emitters. As shown in Table~\ref{tab:ablations} and Fig.~\ref{fig:Ablation}, shrunken emission masks not only cause inaccurate emitters reconstruction, but also damage the overall rendering quality.

\paragraph{Material Recovery}

\begin{figure*}
    \centering
    \includegraphics[width=\textwidth]{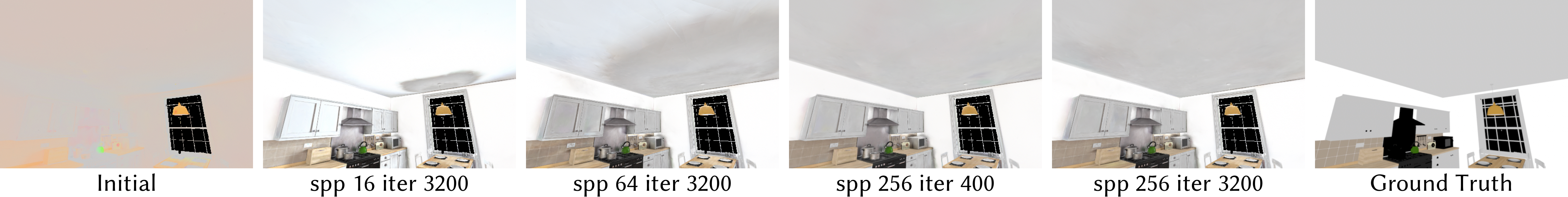}
    \caption{\label{fig:Ablation-material-recovery}
Albedo recovery on \textsc{b-kitchen}.
Low samples per pixel make it difficult to accurately recover the ceiling albedo due to high sampling noise.
    }
    \Description{Ablation-material-recovery}
\end{figure*}

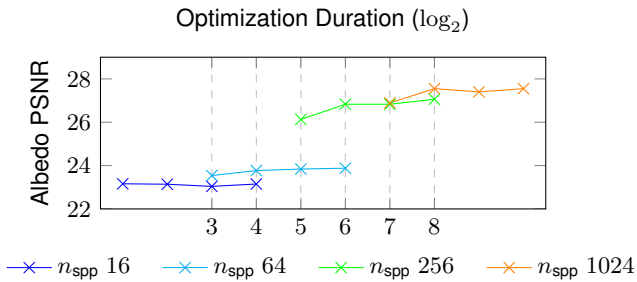
\begin{figure}

\begin{tikzpicture}[font=\small\sffamily]
    \begin{groupplot}[
        group style = {
            group size = 1 by 1,
        },
        height=2.8cm,
        width=0.9\linewidth,
    ]

    \nextgroupplot[
        title = Optimization Duration ($\log_2$),
        ylabel = Albedo PSNR,
        xmin = 0.5,
        xmax = 10.5,
        xtick={3,4,5,6,7,8},
        ymin = 22,
        ymax = 29,
        xmajorgrids = true,
        grid style=dashed,
        legend style={
            draw=none,
            anchor=north,   
            at={(0.5,-0.25)},
            legend columns=4,
            /tikz/every even column/.append style={column sep=1.0em},
        },
    ]
    
    \addplot[color=blue, mark=x, mark size=3pt]
    table[row sep=\\] {
    x  y     \\
    1  23.16 \\
    2  23.14 \\
    3  23.04 \\
    4  23.15 \\
    };
    \addlegendentry{$n_\text{spp} \ 16$}
    
    \addplot[color=cyan, mark=x, mark size=3pt]
    table[row sep=\\] {
    x  y     \\
    3  23.54 \\
    4  23.77 \\
    5  23.84 \\
    6  23.88 \\
    };
    \addlegendentry{$n_\text{spp} \ 64$}

    \addplot[color=green, mark=x, mark size=3pt]
    table[row sep=\\] {
    x  y     \\
    5  26.13 \\
    6  26.83 \\
    7  26.83 \\
    8  27.06 \\
    };
    \addlegendentry{$n_\text{spp} \ 256$}

    \addplot[color=orange, mark=x, mark size=3pt]
    table[row sep=\\] {
    x  y     \\
    7  26.89 \\
    8  27.55 \\
    9  27.40 \\
    10 27.55 \\
    };
    \addlegendentry{$n_\text{spp} \ 1024$}

    \end{groupplot}
\end{tikzpicture}

\caption{\label{fig:graph-material-recovery-albedo-psnr}
    Albedo PSNR during material recovery on \textsc{b-kitchen}: for a fixed optimization budget, higher $n_\text{spp}$ yields higher PSNR, showing that reducing sampling noise is more effective than increasing iterations at low $n_\text{spp}$.
}
\Description{graph-material-recovery-albedo-psnr}
\end{figure}

Results in Fig.~\ref{fig:Ablation-material-recovery} and Fig.~\ref{fig:graph-material-recovery-albedo-psnr} show the effectiveness of diffuse material recovery during Stage~1.
Based on 1-bounce and differentiable rendering, we can recover the diffuse material well, comparing with albedo ground truths on synthetic scene \textsc{b-kitchen}.
Another finding is that, when using the same durations to optimize material, higher $n_\text{spp}$s ($n_\text{spp}$ 256 iter 400) can yield better results than lower $n_\text{spp}$s ($n_\text{spp}$ 64 iter 1600) due to lower sampling noise, which is different from the strategy of I2SDF~\cite{I2SDF} that sets $n_\text{spp}$ to 16 and optimizes for 100k iterations over days.

\paragraph{Path Tracing}

We set the bounce limit $\tau_b$ to 7 and $n_\text{spp}$ to 1024 in our path-traced renderings to balance visual quality and computation time. As shown in Fig.~\ref{fig:Ablation}, using a smaller bounce limit (e.g. 3) introduces significant bias and produces overly dark images, while reducing $n_\text{spp}$ (e.g. 256) leads to loss of detail and noticeably blurrier results. Increasing either the bounce limit or $n_\text{spp}$ further improves image quality but also increases rendering time, as reported in Table~\ref{tab:ablations}. In addition, we conduct an experiment demonstrating that path tracing on fewer 2D Gaussians can be faster; details are provided in Appendix~\ref{sec:appendix-discussion}. Finally, unlike radiant scene rendering, path-traced results require a denoiser to obtain clean images, as illustrated in Fig.~\ref{fig:Ablation}.

\section{CONCLUSION}\label{sec:conclusion}

In this work, we propose Emission-Aware Gaussians and Path Tracing (EAG-PT) to introduce correct light transport into previous radiance field reconstruction work.
The core of our method lies in the separation of emitters and non-emitters, differentiable rendering that recovers radiance and diffuse material, and multi-bounce path tracing for final rendering.
Based on the unified representation, 2D Gaussians, that supports ray tracing and light bouncing, we formulate diffuse indoor scene reconstruction and rendering into an integral framework.
This derives much more natural renders in reconstructed indoor scenes after editing, and even better visual quality than mesh-based baseline, which is promising for practical and realistic real-to-sim reconstruction.

However, our method has limitations.
Emission masks for some real-world indoor scenes are complicated and depend on labeling. An automatic way to obtain them should reduce data-processing time.
Our current material model assumes diffuse reflectance; extending the framework to more expressive BRDFs would enhance realism.
Incorporating multiple importance sampling with emitter importance sampling could reduce variance and accelerate convergence during optimization and rendering. Combining with level-of-detail and real-time global illumination techniques may further improve the efficiency of path tracing.
We leave addressing these limitations and exploring possible improvements to future work.

\begin{acks}

This work was partially supported by
Shanghai Artificial Intelligence Laboratory,
the National Natural Science Foundation of China (Grant No. 62502247),
and
the HKU Startup Fund and Institute of Data Science.
We would like to thank the anonymous reviewers for their valuable feedback.

\end{acks}

\newpage
\bibliographystyle{ACM-Reference-Format}
\bibliography{references}

\clearpage
\appendix

The appendices provide supplementary material to support and extend the main paper.
Additional related work is discussed in Appendix~\ref{sec:appendix-additional-related-work}.
Details on deriving emission masks are presented in Appendix~\ref{sec:appendix-emission-masks}.
The image capture pipeline for acquiring linear radiance is described in Appendix~\ref{sec:appendix-capture}.
Further analysis of our method is provided in Appendix~\ref{sec:appendix-discussion}, and potential downstream applications are illustrated in Appendix~\ref{sec:appendix-possible-applications}.

\section{Additional Related Work}\label{sec:appendix-additional-related-work}

\paragraph{Scene Editing in 3DGRT}

Methods such as 3DGRT~\cite{3DGRT} can accomplish scene editing and re-rendering to some extent. It is able to insert reflective mesh objects into the scene and achieve realistic renders.
Yet, the reflection only happens on the newly inserted mesh objects, which is a one-way bounce between the object and scene, instead of bouncing around inside the reconstructed scene.
The way 3DGRT creates shadows is to dim the color of the corresponding radiant 3D Gaussians, similar to the harmonization method used in MV-CoLight~\cite{MV-CoLight}, which is a heuristic approximation instead of a physically-correct calculation.

\paragraph{Path Tracing in EGR}

Though EGR~\cite{EGR} also uses the phrase \emph{path tracing} in their paper, they ignore true light sources in the scene and only refer to the multi-bounce light transport. The main target of EGR is to reconstruct the radiance field for reflection (like environment modeling in PBIR-NeRF~\cite{PBIR-NeRF}), and they do not optimize material through inverse rendering. While our main target is to recover the SVBRDF properties, and to support editing and re-rendering of the whole scene.

\paragraph{Volumetric Path Tracing}

There are some works using \emph{volumetric path tracing} based on 3D Gaussians.
DSurG~\cite{DSurG} focuses on modeling radiant semi-transparent media, which is not related with our goal. We only use the volumetric representation for the scene, while still use the general path tracing to bounce light when "hitting" the surface.
UGP~\cite{UGP} tends to use a modified 3D Gaussians as a general scene representation. However, such representation is rather complicated for indoor scene reconstruction, and it is mainly designed for appearance modeling and forward rendering; the exploration of inverse rendering on such new representation is still preliminary.

\paragraph{Single Image Relighting}

We are happy to see lots of 2D relighting methods~\cite{LightLab} that achieve photo-realistic scene editing results.
However, it is difficult for these methods to achieve 3D consistent results.
For example, changing the light in one image does not affect another image in the same scene.
\cite{PSDR-Room,S25-Poster-InteractiveObjectInsertion} start from single image yet adopt explicit 3D representation to solve 3D consistency issue, though a single image only has small scene coverage. This does not satisfy interactive roaming in 3D scene.

\paragraph{Asset Retrieval}

Other systems~\cite{PSDR-Room,LiteReality} retrieve and adjust CG assets from input images, but the resulting scenes are typically less photo-realistic than the captured data.

\section{Emission Mask Derivation}\label{sec:appendix-emission-masks}

\begin{figure}
    \centering
    \includegraphics[width=\linewidth]{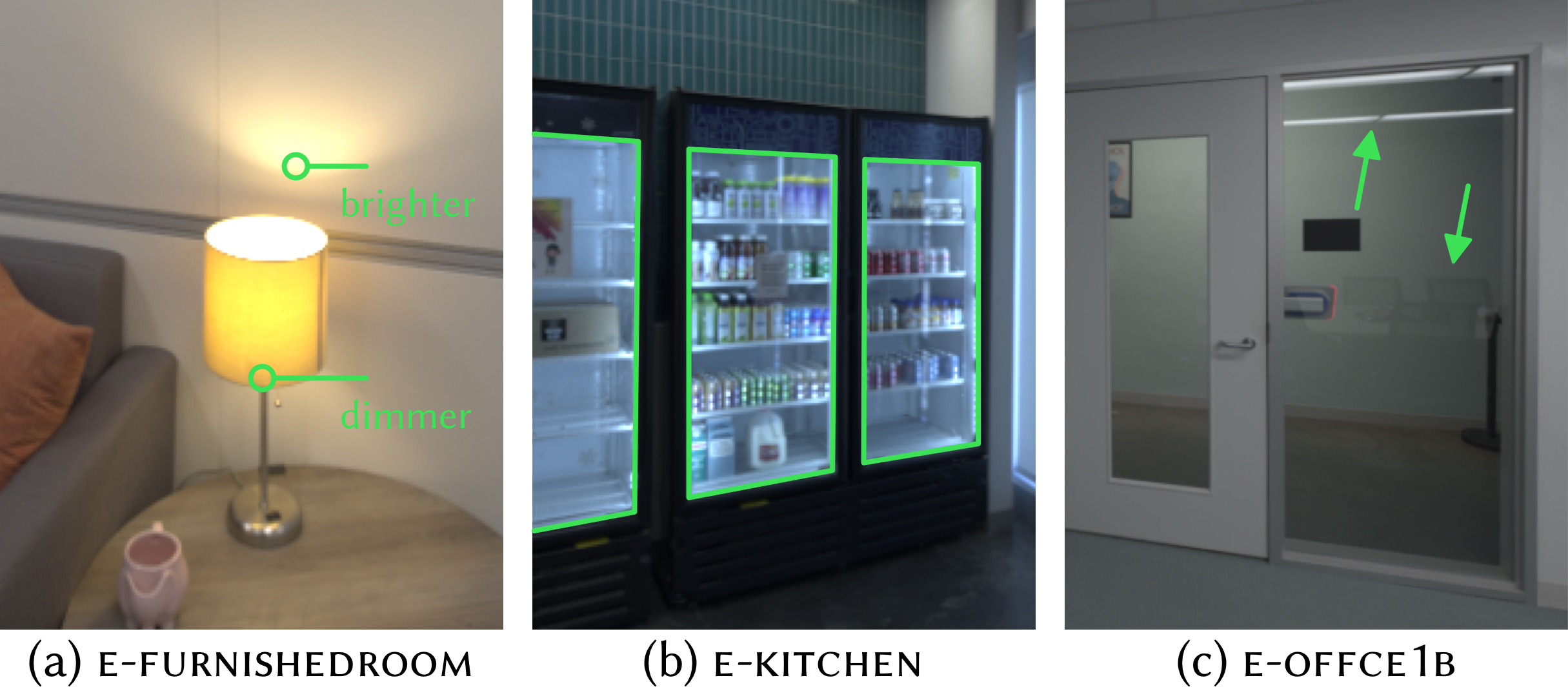}
    \caption{\label{fig:Appendix-emission-mask}
        Exceptions of emission masks.
        (a) Reflection is brighter than emitter.
        (b) Human-defined emitters.
        (c) Emitter with reflection.
    }
    \Description{Appendix-emission-mask}
\end{figure}

Currently, our method relies on 2D emission masks to separate emitters from non-emitters.
For images in linear radiance, emission masks can typically be obtained by simple
thresholding, since emitters exhibit high radiance.
For most scenes, including \textsc{b-}, \textsc{f-}, and \textsc{lectureroom},
we classify a pixel as emissive if its radiance exceeds a scene-dependent
threshold $\tau_R \in \{1.0, 1.5, 2.0\}$.
This method is similar to IRIS~\cite{IRIS}, which applies a threshold of $0.99$ to SDR images.

However, thresholding alone is insufficient in several cases.
First, strong reflections can be brighter than the actual light sources, causing reflective surfaces to be mislabeled as emitters, while genuinely dimmer emitters may fall below the threshold, as illustrated in Fig.~\ref{fig:Appendix-emission-mask}~(a).
Second, emission may be defined semantically: for example, bottles inside the vending machine can be treated as emitters when they are unimportant for subsequent scene editing, as shown in Fig.~\ref{fig:Appendix-emission-mask}~(b).
To handle such cases, we manually refine emission masks with SAM~\cite{SAM} for efficient annotation.
The final emission mask is the union of the thresholded and manually labeled regions, $M = M_{\text{threshold}} \cup M_{\text{SAM}}$.
One point worth noting is that emission masks are not necessary for every input image.
For three scenes in \textsc{e-}, each containing thousands of multi-view training images, we annotated emission masks for only a small subset (1/19, from a single camera), and set $\mathcal{L}_e = 0$ in Eq.~\ref{eq:optimization-stage-0} for the unlabeled images.
Despite this sparse supervision, our method still reliably identified emitters.
To reduce the preprocessing cost for newly captured indoor scenes, a promising direction would be automatic emitter annotation using a computer vision detector.
Beyond 2D labeling, emitters could also be annotated directly in 3D after radiant scene reconstruction, either manually or through point-cloud classification.

Besides, the use of emission masks assumes, as in Sec.~\ref{sec:method-material-recovery}, that emitters do not reflect light.
Real indoor scenes can violate this assumption: for instance, the glass door in \textsc{e-office1b} should be treated as an emitter but is also highly reflective, as shown in Fig.~\ref{fig:Appendix-emission-mask}~(c).
Currently, our method does not explicitly model such mixed emitter-reflector materials.

\newpage

\section{Real-World Scene Capture}\label{sec:appendix-capture}

Our method operates on calibrated multi-view images in linear radiance space. Following VR-NeRF~\cite{VR-NeRF} and FIPT~\cite{FIPT}, we capture the indoor scene \textsc{lectureroom} with a SONY ZV-E10M2 APS-C camera equipped with the default kit lens and mounted on a tripod. The camera is operated in full manual mode with aperture fixed to f/8.0, ISO to 100, and focal length to 16\,mm to obtain a wide field of view and a stable radiometric response. Before capture, we set the white balance using a gray card placed in the scene and then fix it for the entire sequence to ensure consistent color calibration across all views. The lens is switched to manual focus with the focus distance set to 1.0\,m, and kept unchanged during capture to avoid per-view focus variations.
To avoid clipping in shadows and highlights, we determine a bracketing range of exposure times in the scene such that the shortest exposure resolves the brightest emitters and the longest exposure resolves the darkest shadows. We then interpolate within this range to acquire multiple exposures per viewpoint. We rotate the camera, adjust the tripod height, and translate the tripod to obtain dense multi-view coverage of the scene.

After capture, for each viewpoint, the captured RAW images are converted to linear radiance in the range $[0,1]$ and merged into a single linear image. We then apply lens undistortion and vignetting correction to every image. We run COLMAP on sRGB images converted from linear images to recover intrinsics and extrinsics. The recovered poses are subsequently rotated so that the scene floor aligns with the $+z$ axis of the world coordinate system, which simplifies downstream editing and rendering. This pipeline is sufficient to produce input data for our reconstruction method from generic indoor scenes.

However, capturing a scene under a single lighting condition is insufficient for evaluating real-scene relighting performance. While one-light-at-a-time (OLAT) datasets have been studied for real-world objects~\cite{DeferredNeuralLighting,ZhiyiOLAT}, there are very few real-world indoor multi-view datasets with multiple controllable light configurations. FIPT~\cite{FIPT} shows several relit images in their paper but does not release the full dataset.
We regard multi-condition, multi-view indoor data as an important resource for physically based indoor reconstruction and relighting research. As an initial step, we capture \textsc{lectureroom} under three distinct lighting configurations: all lights on, only the front-half lights on, and only the back-half lights on. At each camera position, we keep the camera rigidly fixed and switch the scene among these three light conditions; for each condition, we record a multi-exposure bracket as described above. For \textsc{lectureroom}, we capture 100 distinct camera positions, each with 3 lighting conditions. We show representative ground-truth images in Fig.~\ref{fig:Result-real-lectureroom-relighting}, together with path-traced renderings produced by our method, for evaluating the relighting capability of our method on real scene. We will release this multi-condition \textsc{lectureroom} dataset to the research community.

Looking forward, extending this capture process to a broader variety of indoor environments and to a richer set of lighting conditions would provide valuable benchmarks for disentangling geometry, materials, and illumination, and for advancing inverse rendering and indoor scene relighting methods.

\section{Discussion}\label{sec:appendix-discussion}

\paragraph{Method Input}

Our method currently requires multi-view images with linear radiance as input. Although the capture pipeline described in Appendix~\ref{sec:appendix-capture} is physically faithful, it remains  time-consuming for typical users.
A promising direction is to integrate techniques for camera response function recovery and HDR restoration~\cite{IRIS,PPISP}, which could relax this requirement and allow more convenient SDR inputs, albeit with some degradation in visual fidelity. For example, saturated regions such as the clipped ceiling light are unlikely to be fully recovered.

At present, our method optimizes all scenes using $540\times360$ input images. While increasing the input resolution is a straightforward way to improve reconstruction quality, further algorithmic and systems-level optimization would be valuable to reduce the overall processing time.

Additionally, EAG-PT assumes multi-view capture that observes most of the scene, as described in Sec.~\ref{sec:method}. In practice, however, it is impossible to sample all points in 3D space. Occluded or rarely seen regions can acquire an incorrect radiance cache in Stage~0, which in turn degrades albedo recovery in Stage~1. As presented in Fig.~\ref{fig:Method-pipeline}, albedo of the ceiling above the projector appears darker than expected due to insufficient observations. Integrating radiance regularization and optimization strategies as in \cite{FIPT,EGR}, or more generally coupling EAG-PT with priors for unobserved regions, could mitigate such artifacts and improve robustness under incomplete coverage.

\paragraph{Modeling}

In our current formulation, Gaussians are attached with diffuse albedos. While this already yields high-quality reconstructions and renderings, it still falls short of the richness of real-world materials. For example, highly specular objects such as the metallic range hood and microwave oven in Fig.~\ref{fig:graph-material-recovery-albedo-psnr} are not faithfully reproduced. In real scenes (e.g., Figs.~\ref{fig:Result-real-comparison-with-fipt},\ref{fig:Result-real-lectureroom-relighting}), reflections on the whiteboard and ceiling are also difficult to capture. Extending EAG-PT with a compact parametric model, such as a simplified Disney BRDF~\cite{DisneyBRDF}~\cite{GS-IR,GaussianShader,IRGS}, and with refraction modeling, should further improve realism and enable a broader range of appearance effects. Besides, although adding new terms such as roughness and metallic in Eq.~\ref{eq:2d-gaussians-ray-tracing-values} is straightforward, differentiable estimation of these parameters might be difficult.

Moreover, the indoor scenes considered in this work are predominantly confined and dominated by artificial illumination, whereas real indoor environments often receive significant external lighting, as in \textsc{e-officeview1}, \textsc{e-officeview2}, and \textsc{e-riverview}. Incorporating an environment map with explicit window modeling, as in MILO~\cite{FIPT-ref-MILO}, is a promising direction to extend EAG-PT to more diverse and open indoor configurations.

\paragraph{Optimization Strategy}

Several optimization strategies could further improve quality within our current framework.
Because 2D Gaussians are semi-transparent, the reconstructed surface has non-negligible thickness rather than being truly planar, which can cause self-intersection artifacts near surfaces. Regularization terms such as depth distortion in 2DGS~\cite{2DGS} could help alleviate this issue.
In addition, the number of 2D Gaussians is currently fixed throughout optimization. Incorporating adaptive control~\cite{3DGS} would likely remove low-opacity floaters and improve efficiency.
Finally, because our method relies on ray tracing rather than rasterization, the optimizations in Eqs.~\ref{eq:optimization-stage-0},\ref{eq:optimization-stage-1},\ref{eq:optimization-light-baking} could further benefit from stochastic ray sampling across input views~\cite{3DGRT,VR-NeRF}.

\paragraph{Path Tracing}

While our path tracer produces photo-realistic results after editing, the current rendering speed remains far from real-time. The two dominant bottlenecks are repeated ray-Gaussian intersections and the large number of required samples.

\begin{figure}
    \centering
    \includegraphics[width=\linewidth]{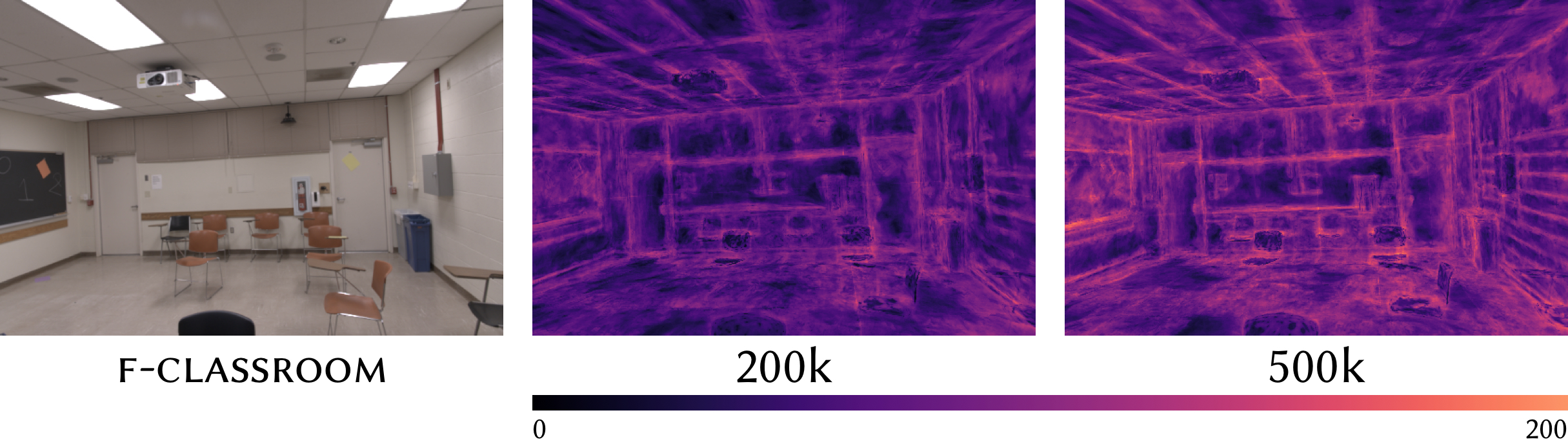}
    \caption{\label{fig:Appendix-hit-count}
        Ray-Gaussian intersection count visualization on two versions of reconstructed \textsc{f-classroom}s with different Gaussian count.
    }
    \Description{Appendix-hit-count}
\end{figure}

We re-trained a variant \textsc{f-classroom} with fewer 2D Gaussians (200k instead of 500k), which reduced the average intersection count per ray from 79 to 63 (visualized in Fig.~\ref{fig:Appendix-hit-count}) and improved rendering speed by approximately 11\%. Instead of retraining separate checkpoints, applying a level-of-detail hierarchy to the original reconstruction~\cite{V3DG} is an attractive alternative for accelerating rendering. In addition, more compact Gaussian primitives~\cite{3DGRT} could further reduce intersection counts and improve performance.

On the sampling side, our current implementation only uses cosine-weighted sampling for secondary rays. Although this reduces noise compared to uniform sampling, it is still insufficient to obtain clean images at low sample counts. Adopting multiple importance sampling with emitter importance sampling targeted at emissive Gaussian primitives, as in PBIR-NeRF~\cite{PBIR-NeRF}, should substantially reduce variance when emitters are tiny and enable fewer samples for smoother path tracing during scene editing.

\paragraph{Instance-Level Reconstruction}

Like most prior indoor inverse rendering methods, ours treats the scene as a single undifferentiated instance, without explicit object- or semantic-level structure.
Consequently, we currently rely on box selection of Gaussians for editing, which often leaves extruding Gaussians at object boundaries, as in Fig.~\ref{fig:Result-synthetic-lightball-relighting}. Similar problems arise when removing objects: newly exposed regions correspond to previously unseen surfaces, degrading realism and geometric consistency.

Incorporating instance-level segmentation~\cite{grounded-sam,object-gs} into EAG-PT would enable object-aware reconstruction and editing, reducing boundary artifacts and simplifying user interaction. Coupling such segmentation with generative 3D completion models~\cite{sam-3d} is another promising avenue to plausibly hallucinate newly visible regions and improve both reconstruction quality and editing flexibility in complex indoor scenes.

\newpage

\section{Possible Applications}\label{sec:appendix-possible-applications}

While our experiments focus on reconstruction and editing quality, we briefly outline two downstream applications that could benefit from EAG-PT. These use cases are illustrative only; we do not conduct task-level evaluations.

\paragraph{Interior Design} A homeowner first captures their existing indoor space, and the captured data are provided to an interior designer. The designer reconstructs the scene with EAG-PT and iteratively explores multiple editing proposals by modifying furniture layout, materials, and lighting. Because EAG-PT preserves realistic, physically consistent illumination after editing, the designer can then light-bake the edited scenes to obtain 3D assets that can be previewed in real time (e.g., on a desktop viewer or XR device), enabling the client to compare design alternatives before committing to a final furnishing plan.

\paragraph{Embodied AI Post-Training} Prior to deploying a robot in a specific real indoor environment, we reconstruct that environment with EAG-PT and generate a family of edited variants that simulate plausible changes in layout, object placement, and lighting conditions. Training or fine-tuning the robot’s perception and control policies in these photorealistic, emission-aware variants can reduce the sim-to-real gap, by exposing the policy to realistic illumination effects (shadows, interreflections, bright emissive sources) and moderate structural changes in advance. This may provide improved robustness to lighting and layout changes encountered during deployment.

\end{document}